\documentclass[aps,prl,twocolumn,superscriptaddress,showpacs]{revtex4}

\usepackage{amsmath,epsfig,color,amssymb}
\usepackage{graphicx}
\usepackage{graphicx}
\usepackage{dcolumn}
\usepackage{bm}
\usepackage{graphicx}
\usepackage{amsmath, amsfonts, amssymb,mathrsfs}
\usepackage{pstricks}
\usepackage{amsxtra}
\usepackage{amsthm}
\usepackage{natbib}

\makeatletter

\def\be{\begin{equation}}
\def\ee{\end{equation}}
\def\bea{\begin{eqnarray}}
\def\eea{\end{eqnarray}}
\newcommand{\ket}[1]{\mbox{$|#1\rangle$}}
\newcommand{\bra}[1]{\mbox{$\langle#1|$}}

\def\be{\begin{equation}}      
\def\ee{\end{equation}}
\def\beu{\begin{equation*}}   
\def\eeu{\end{equation*}}

\providecommand{\abs}[1]{\left\lvert#1\right\rvert}   
\providecommand{\ket}[1]{\left|#1\right\rangle}

\providecommand{\bra}[1]{\left\langle#1\right|}

\providecommand{\comm}[2]{\left[ #1, #2 \right]}  		

\graphicspath{../}

\begin{document}

\title{A Quantum Network of Silicon Qubits using Mid-Infrared Graphene Plasmons}
\author{M.~J.~Gullans}
\email[Corresponding author: michael.gullans@nist.gov]{}
\affiliation{Joint Quantum Institute, National Institute of Standards and Technology, Gaithersburg, MD 20899, USA}
\author{J.~M.~Taylor}
\affiliation{Joint Quantum Institute, National Institute of Standards and Technology, Gaithersburg, MD 20899, USA}
\date{\today}


\begin{abstract}
We consider a quantum network of mid-infrared, graphene plasmons coupled to the hydrogen-like excited states of group-V donors in silicon.  First, we show how to use plasmon-enhanced light-matter interactions to achieve single-shot spin readout of the donor qubits via optical excitation and electrical detection of the emitted plasmons.    We then show how plasmons in high mobility graphene nanoribbons can be used to achieve high-fidelity, two-qubit gates and entanglement of  distant Si donor qubits.  The proposed device is readily compatible with existing technology and fabrication methods.
\end{abstract}
\maketitle

The spins of Group-V donors (P, Bi, As, Sb) in silicon have long been intriguing as potential quantum bits \cite{Zwanenburg13}.  Dramatic advances in the implementation of these ideas continue apace, including demonstration of state initialization and readout, robust single qubit operations \cite{Zwanenburg13,Yang06,Yang09,Sekiguchi10,Wolfowicz13}, and quantum memory exceeding an hour \cite{Saeedi13}.  However, localized control, single-shot readout \cite{Morello10,Pla12}, and coherent coupling between two qubits \cite{Koiller01,Weber14} remain challenging.  One approach to overcome these challenges, which has been successful in other solid-state impurity systems such as nitrogen-vacancy centers in diamond \cite{Jelezko06} and quantum dots \cite{Hennenberger08}, is to use higher orbital states to control the spin optically \cite{Vinh08,Greenland10}.  This allows mapping of the spin-state into light, which can be used to both readout the spin and induce entanglement between  distant impurities \cite{Bernien13}.  
In the case of the group-V donors in Si, such higher orbital excitations lie in the mid- to far-infrared, making it challenging to use conventional optics to achieve this goal.  

Recently, graphene has  emerged as a powerful platform for plasmonics in this mid-infrared frequency regime \cite{Jablan09,GarciadeAbajo14}.  Graphene's high mobility \cite{Mayorov11,Elias11,Baringhaus14}, versatile fabrication, and the ability to tune the plasmon properties via external gate voltages has already suggested many potential applications in this challenging portion of the electromagnetic spectrum \cite{Low14}.  Furthermore, Si is a natural  base substrate for graphene plasmonics due to its ubiquitousness in fabrication technology and its low loss in the mid-infrared.  This coincidence of properties suggests that graphene plasmons may be a powerful resource for orbital control of Si impurity qubits.

In this Letter, we propose to achieve this control by coupling the Si impurities to mid-infrared  plasmons in a graphene nanoribbon.
  First, we show how to use the plasmon enhanced light-matter interaction to achieve single-shot spin readout.    We then show how plasmons in high mobility nanoribbons can be used to achieve high-fidelity, two-qubit gates and entanglement of  distant Si donor qubits.

\begin{figure}[h]
\begin{center}
\includegraphics[width=8 cm]{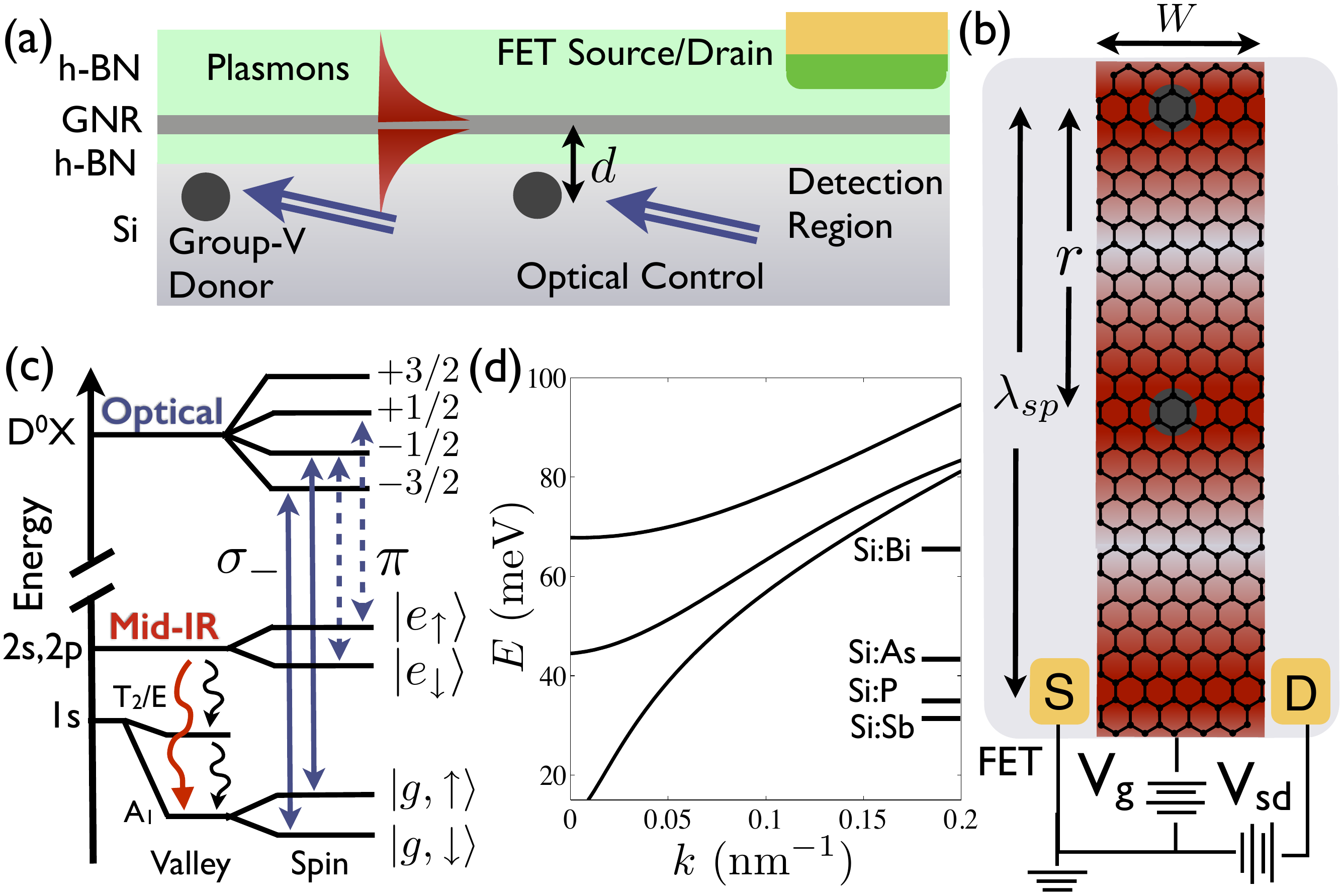}
\label{fig:SiVfig1}
\caption{(a) Two Si dopant qubits are coupled via plasmons in a graphene nanoribbon (GNR) that is encapsulated in hexagonal-BN (h-BN).  The plasmons are detected by a field effect transitor (FET), allowing single-shot readout of the spin state.  $d$ is the distance from the qubit to the nanoribbon. (b) Top view of the device.  The color scale shows the electric field intensity of the plasmon mode.  $W$ is the width of the nano ribbon, $\lambda_{sp}$ is the wavelength of the plasmon, and $r=n\, \lambda_{sp}/2$ ($n$=1 for the case shown) is the qubit separation. (c) Spectrum of the group-V donors in Si with the pathway for the plasmon decay (red) and phonon decay (black).  We assume a strain is applied to make the $2s$ and $2p$ state degenerate, which allows spin selective excitation from the 1$s$ ground states to the $2p$ states through the exciton state $D^0X$ by controlling the polarization of the excitation light.  We work with the reduced set of states: the two electron spin ground state labeled $\ket{g,s}$ and a Kramer's pair in the 2s-2p manifold $\ket{e_s}$.
(d) Dispersion of the first three plasmon modes in the nanoribbon ($W=50$ nm and $n_e=10^{12}$ cm$^{-2}$) and Si:Sb, P, As 1$s$ ground to $2p_0$ transitions and Si:Bi 1$s$ to $2p_\pm$ transition.
}
\end{center}
\end{figure}

The device we consider is illustrated in Fig.\ 1(a-b).  Shallow dopant atoms in Si are coupled via plasmons in a graphene nanoribbon (GNR).   
The excited states of the impurity have a hydrogen-like spectrum with a manifold of $1s$  ground states and excited $s,~p,~\ldots$ states [Fig.\ 1(c)] \cite{Ramdas81}.  As illustrated in Fig.\ 1(d), these transitions have a strong overlap with the plasmonic modes of the nanoribbons.  This gives rise to a strong Purcell enhancement of the radiation from the excited states into plasmons  \cite{Chang07,Koppens11}.  This is the crucial feature of this system, which, when combined with spin-selective excitation, allows control, readout and entanglement of the impurity spins.

In our analysis, we restrict ourselves to the reduced level-scheme labeled in Fig.\ 1(c) and detailed in the supplemental material \cite{supp}.   There are two electron spin ground states $\ket{g,s}$ [$1s(A_1)$ state] coupled via optical, spin selective excitation to a Kramer's doublet of excited states $\ket{e_s}$ (hybridized $2p$-$2s$ state).  The excited state can decay back to the ground state either through phonons or plasmons.
 Spin-selective excitation is achieved with two methods that take advantage of optical excitation to the spin-3/2, donor bound, exciton state $D^0X$ as the first leg in a two photon transition from the $1s$ ground state to the $2p$ states.   The first, demonstrated in Fig.\ 1(c), uses circularly polarized light on one leg of the lambda transition and linearly polarized light on the second. We refer to this as the $\sigma_\pm \pi$-excitation scheme.  This will only excite one spin state, e.g., spin-down to spin-up as in Fig.\ 1(c), while leaving the other spin state unaffected.  In the second approach, we apply opposite circular polarization on each leg, which excites both spin states, but at different rates due to the difference in Clebsch-Gordon coefficients.  We refer to this as the $\sigma_\mp \sigma_\pm$-excitation scheme. 
  
We first show how the strong Purcell enhancement can be combined with electrical detection to achieve single-shot readout.
Plasmonic excitations in graphene nanoribbons have been systematically characterized in Ref.~\cite{Christensen12}.
Their properties are similar to optical waveguides with a discrete spectrum of transverse modes that reduces to a single mode as the width decreases to zero.
In the single mode regime of a nanoribbon with width $W$ and length $L$, the emission rate into the plasmons can be approximated by Fermi's golden rule $\gamma_p = 2 \pi g^2_{k}(\omega) D(\omega)$ where $D(\omega) = L/2 \pi v_g$ is the density of states and $v_g = d\omega /dk = \omega W \abs{\eta'}/2$ is the group velocity.  Here $\eta'$ (dimensionless and $\sim 1$) is the derivative of a universal function $\eta(k W)$  associated with the lowest order mode  (defined below) and $g_k$, analyzed in detail later, can be interpreted as the quantized electric field per plasmon times the dipole moment of the impurity.   The small mode volume and reduced group velocity of the plasmon compared to free space gives the Purcell enhancement for a single impurity 
\be\label{eqn:purcell}
\frac{\gamma_p}{\gamma_r} = \frac{3\, \chi}{16} \frac{c^3}{v_g^3}\,\abs{\eta'(k W)}^2 \, e^{-4 \pi d/\lambda_{sp}},
\ee
where $d$ is the distance from the impurity to the nanoribbon, $\chi$ is a factor due to the surrounding dielectric environment, $c$ is the speed of light, $\gamma_r \sim 1$ kHz  is the radiative decay rate of the $e_s$ states, and $\lambda_{sp}=2\pi/k$ is the plasmon wavelength.  As $v_g$ approaches speeds as low as the Fermi velocity $v_F\approx 10^6$ m/s, the Purcell enhancement  approaches $10^8$ compared to free space \cite{Koppens11}.  The  enhancement  relative to the phonon broadening $\gamma$ is shown for the Si:P $1s(A_1)-2p_0$ transition in Fig.\ 2(a) and for the Si:Bi $1s(A_1)$-2$p_\pm$ transition.   Crucially, near the nanoribbon, the plasmon emission exceeds the phonon decay by several orders of magnitude, enabling deterministic conversion of the excited state population into a propagating plasmon in the nanoribbon.

To achieve single shot-readout we use the $\sigma_- \sigma_+$-excitation scheme.  In this case, the spin-down state will emit plasmons at nine times the rate of the spin-up state, enabling one to distinguish the spin-states via the difference in the number of emitted plasmons.
However,  when the excited state relaxes via phonon decay, the spin will become depolarized by the 1$s(T_2/E)$ valley states which have a large spin-orbit coupling.  As a result, the optimal excitation time scales as $\sqrt{\gamma/\gamma_p}$ and, thus, the excess in the emitted plasmon number will scale with the same factor \cite{supp}.  For large Purcell enhancement, one could achieve single-shot readout of the impurity spin by detecting the plasmons electrically with a field effect transistor (FET) [Fig.\ 1(b)] as was demonstrated for optical plasmons in Ref.\ \cite{Falk09}.   Figure\ 2(b) shows the  distribution of the emitted plasmon number for $\gamma_p/\gamma=300$, where the two distributions can be distinguished with a  fidelity of $96\%$, demonstrating the potential for single-shot readout.

\begin{figure}[t]
\begin{center}
\includegraphics[width=8.46 cm]{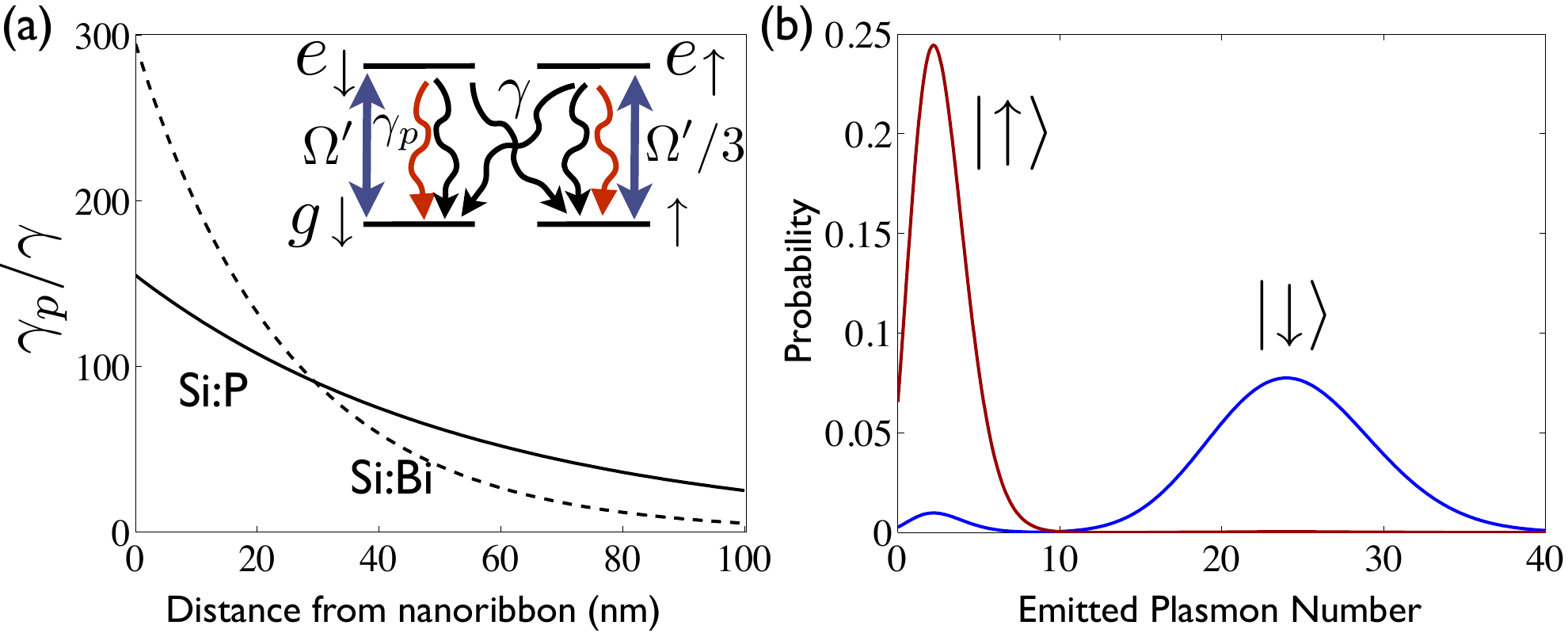}
\label{fig:SiVfig2}
\caption{(a) Enhancement of plasmon decay compared to phonon decay with vertical distance of the qubit from the nanoribbon for Si:P/Bi (solid/dashed).  Here $W=10/5$ nm  and $n_e=2 \cdot 10^{11}/7\cdot 10^{11}$ cm$^{-2}$ for P/Bi. (inset) Reduced level diagram with the $\sigma_-\sigma_+$-excitation scheme.  Plasmon decay (red) occurs at an enhanced rate compared to phonon decay (black). (b) Finite time fluorescence distribution conditioned on the initial spin state  (here $\gamma_p/\gamma=300$).   Readout fidelity is determined by the distinguishability of the two distributions.  In the case shown the spin-up state (red) can be distinguished from the spin-down state (blue) with a fidelity as high as 96$\%$ by assigning events with more than 10 detected plasmons as spin-down.}
\end{center}
\end{figure}

 We have now established the necessary ingredients to build a quantum network of the impurities.  However, before we can give a rigorous analysis of impurity entanglement and two-qubit gates, we need to explore the details of the plasmonic modes and the spin-impurity coupling.
  
 Focusing on these details, we define the universal function $\eta$ for the plasmons \cite{Christensen12}.  An important difference from conventional optics is that the plasmons are near field excitations and can be accurately described by electrostatics.  As a result, the free space wavelength does not set an absolute scale and the dispersion relation $\omega_j(k)$ of each transverse mode $j$ is determined by the width of the nanoribbon $W$ and the universal function $\eta_j(x)$ 
\be \label{eqn:eta}
\omega_j(k)= \frac{\chi}{4 \pi \epsilon_0} \frac{\mathrm{Im} [ \sigma (\omega_j)]}{\eta_j(k W) \, W}.
\ee
Here $\epsilon_0$ is the dielectric constant and the conductivity is well approximated by the Drude formula $\sigma(\omega)= \frac{i e^2}{\pi \hbar} \frac{\omega_F}{\omega+i \kappa}$ for frequencies far below twice the Fermi frequency $\omega_F$ ($\kappa$ is the inverse scattering time, $\hbar$ is Planck's constant and $e$ is the elementary charge) \cite{Thongrattanasiri12}.  In Fig.\ 1(d) we show the dispersion for the first three modes using the universal functions found in Ref.~\cite{Christensen12}.

The coupling $g_k$ between a plasmon and the orbital state of the impurity is determined by the interaction Hamiltonian $H_I=-\bm{\mu}_I\cdot \bm{E}$, where $\bm{\mu}_I$ is the dipole moment of the impurity and $\bm{E}$ is the electric field of the graphene plasmon.  For small enough widths that the impurity is only resonant with the first mode in Fig.\ 1(d), we can write the Hamiltonian for a pair of dopant atoms interacting with the plasmons as
\begin{align} \nonumber
H&= \omega_d \sum_{i,s} \ket{e_{is}}\bra{e_{is}} 
+ \sum_k\omega(k) a_k^\dagger a_k \\ 
&+\sum_{k,s} g_{k} a_{k} (\sigma_{1s}^+ +e^{i k r} \,\sigma_{2s}^+ ) + h.c.,
\end{align}
where we take the dipole polarized perpendicular to the nanoribbon.  Here $\sigma_{is}^+=\ket{e_{is}}\bra{g_i,s}$, $\sigma_{is}^-=(\sigma_{is}^+)^\dagger$, $a_k$ are the bosonic operators for the plasmon modes, $\omega_d$ is the transition frequency of the donor, $\omega_k$ is the plasmon frequency, and $g_k$ is plasmon-impurity coupling.  The lowest order mode is a monopole with the charge density approximately constant across the nanoribbon.  Knowing the mode function and the dispersion allows us to derive an analytical expression for $g_k= \frac{e\, \chi \mu_I}{4 \pi \epsilon_0 \eta W}\sqrt{\frac{\omega_F}{\omega(k)\pi L W}}e^{-k z}$ by quantizing the plasmon energy, which has contributions from both kinetic and electromagnetic energy  \cite{Gullans13}.

 The system we have described falls into a well  known class of quantum optics models where several emitters are strongly coupled to a common, infinite range (without losses) optical mode.  It is analogous to a cavity quantum electrodynamics (cQED) model in the weak coupling, but large Purcell factor limit.  As a result, we can draw on several established techniques for achieving two-qubit gates \cite{Pellizzari95,Domokos95} and entanglement \cite{Reiter12} in these models.  The enabling feature is the emergence of super-radiance due to quantum interference of the emitted light into a single mode \cite{Gross82}.  In particular, when $ kr = n \pi$ for integers $n$ the plasmon emission from the two impurities maximally interferes and the system is super-radiant because there are two states $\ket{ee}$ and $\ket{ge} + (-1)^n \ket{ge}$ with collectively enhanced decay $2 \gamma_p $ and one sub-radiant state $\ket{ge} - (-1)^n \ket{eg}$, which decays at the rate $\delta \gamma=\gamma+\gamma_p\, \big(1-e^{-r/L_p}\big)$, where $r$ is the distance between the impurities and $L_p = 2 \mu\,v_g \omega_F/e\,  v_F^2$ is the plasmon propagation length, which increases with the mobility $\mu$ \cite{Jablan09}.  Room temperature mobilities approaching 10$^6$ cm$^2$/Vs have been demonstrated for bulk graphene surrounded by h-BN \cite{Mayorov11,Elias11}, which would allow mid-infrared plasmons to coherently propagate for several microns.  The narrowing of the sub-radiant state is measured by the ratio $f=\delta \gamma/\gamma_p$.   
 Below we show how to utilize the sub-radiant state to achieve a two-qubit phase gate and entanglement via dissipation.  For these protocols, the error probability scales with $f$, making it crucially important that $f\ll 1$.  Figure~3(a) shows $f$ for $n=1$ with increasing mobility, where we see that it can be below $1\%$. 
 
  \begin{figure}[t]
\begin{center}
\includegraphics[width=.5 \textwidth]{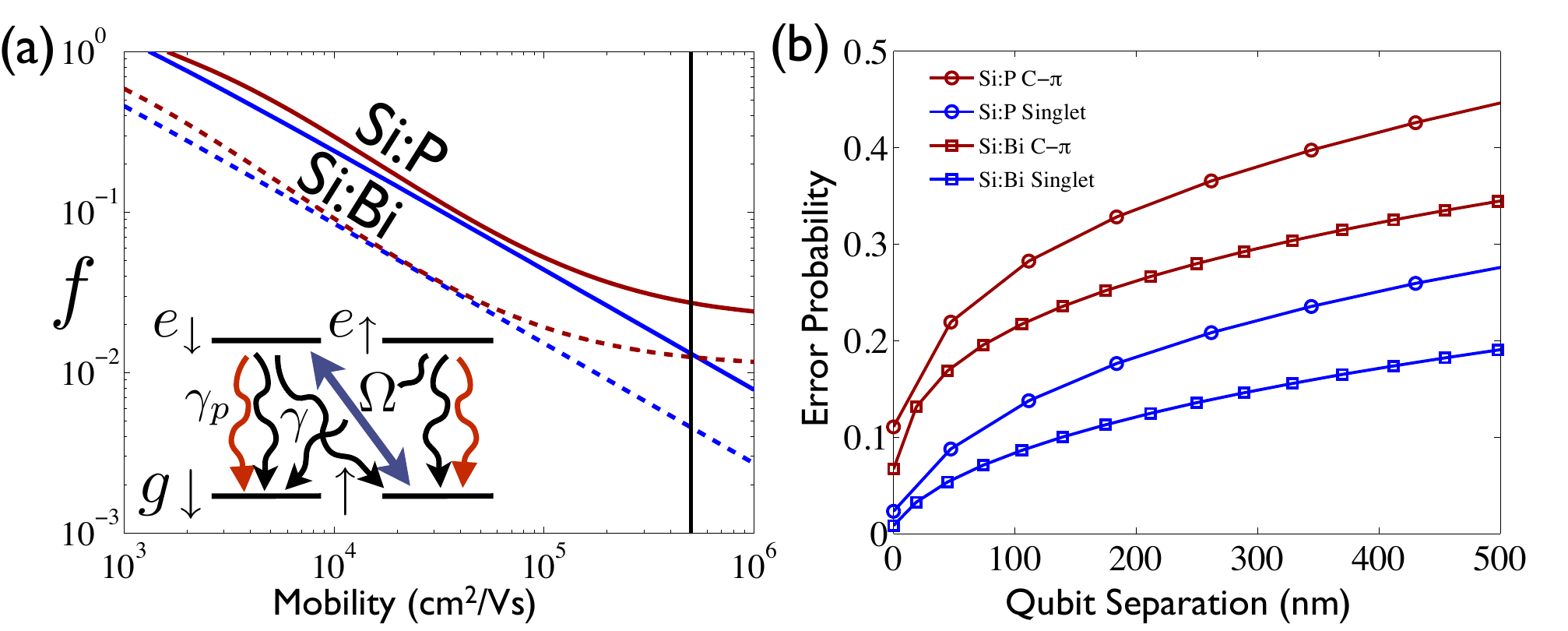}
\label{default}
\caption{
(a) Scaling of the sub radiance narrowing $f$ for $n$=1 with increasing mobility for (red) a fixed width W=20(10) nm and a carrier doping of $n_e=$2$\cdot10^{12}$(7$\cdot10^{12}$) cm$^{-2}$ for P(Bi) and (blue) for an optimized nanoribbon width and carrier doping.  For the optimized case $W$ decreases with the mobility from 25 nm to 3 nm and the doping decreases from $10^{13}$-10$^{11}$ cm$^{-2}$.  The vertical line is the highest observed mobility for graphene encapsulated in h-BN.   (inset) Reduced level diagram with the $\sigma_-\pi$-excitation scheme.
(b) Error probability for two-qubit entanglement preparation and the C-$\pi$ gate for Si:P and Si:Bi.  We take a graphene mobility of $6 \cdot 10^4$ cm$^2/$V$\cdot s$ and optimize  the doping and nanoribbon width at each point in the same range as in (a).}
\end{center}
\end{figure}

 To implement a C-Phase gate we use the $\sigma_-\sigma_+$-excitation scheme, where the driving field takes the form
\begin{align}
H_{c}&=\frac{2\Omega'(t)}{3}\sum_{s}\big(\sigma_{1s}^+ - e^{i \theta}\,  \sigma_{2s}^+ +h.c.) (1-s)
\end{align}
here $s=\pm1/2$, $\Omega'(t)$ is the Raman Rabi frequency of the two-photon transition from $g$ to $e$ and $e^{i \theta}$ is the relative phase of the drive, which we set to $ \textrm{sign} (\cos kr)=(-1)^n$.

All of the two-qubit states are excited to sub-radiant states by this drive, but each at different rates because of a combination spin-selective excitation and the inability to excite the super-radiant states due to a strong quantum Zeno effect.   For a far off resonant, weak drive we can imprint a different phase on each state with minimal dephasing.  The resulting unitary operation is locally equivalent to a C-Phase gate with the phase $e^{i \phi(t)}$ where $\phi(t)= 4 \int dt' \abs{\Omega'(t')}^2/9\delta$ \cite{Pellizzari95,Domokos95}.  Such a gate is universal for quantum computation.
To prevent errors during the gate operation, we require $\delta \gg \delta \gamma $ to prevent dephasing from the sub-radiant state.  In addition, we require $\delta \ll \gamma_c$ because the phase picked up from the super-radiant states cancels out the desired phase on $\ket{\uparrow \uparrow}$.  These two constraints lead to the optimal, average gate fidelity when $\delta \sim \sqrt{\delta \gamma\, \gamma_c}$  as
$
\mathcal{F}_\phi \approx 1 - 3 \sqrt{3}\, \sqrt{{f}} \, \phi/4
$
Fig.\ 3(b) shows the minimized error probability $1-\mathcal{F}_\pi$ for a C-$\pi$ gate with increasing dopant separation.  

The two-qubit gate described above is coherent and deterministic; as a result, any dephasing from phonons or plasmon loss will lead to errors.  To create an entangled state between the two qubits, one can overcome this limitation by engineering the desired state to be a unique dark state of some driven, open system dynamics.  For our system, where two-qubits are coupled via an excited state to a common bosonic mode, it has been well established that one can drive the qubits towards a maximally entangled state with an error probability that scales linearly in $f$ instead of as the square root \cite{Reiter12}.   We can achieve this in our case with the $\sigma_-\pi$-excitation scheme that excites spin-up electrons to spin-down electrons in the excited state
\be
H_c = \Omega(t)\big(\ket{e_{1\downarrow}}\bra{g_1,\uparrow} - e^{i \theta} \, \ket{e_{2\downarrow}}\bra{g_2,\uparrow}  + h.c. \big)
\ee
where $\Omega$ is the Raman rabi frequency and $\theta$ is defined as above.

With the addition of transverse magnetic field, the singlet state becomes the unique dark state because  it is not coupled to any sub-radiant states or triplet states.   As a result the system is optically pumped into the singlet state.  
We give only a brief description here, since similar schemes based on cQED and plasmons have been extensively analyzed \cite{Reiter12}.  The crucial feature is that when the triplet states are excited to the sub-radiant states, at rate $\Omega^2/\delta \gamma$, the phonon decay occurs predominantly through the excited 1$s(T_2,E)$ states.  These states have a relatively large spin-orbit coupling due to a large overlap with the donor nucleus [$\sim 0.02(1)$ meV in Si:P(Bi) \cite{Ramdas81}], which will mix the singlet and triplet states before the electron decays back to the ground state.  Thus, there is an optical pumping rate into the singlet state at rate $\sim\Omega^2/\delta \gamma$, while the rate out, $\sim f\Omega^2/\delta \gamma$, is much slower.  For large transverse fields, the dynamics result in the fidelity for singlet state preparation $\mathcal{F}_s \approx 1- 3 f$.  In Fig.\ 3(b) we show the error probability  for singlet state preparation $1-\mathcal{F}_s$ with dopant separation, where it goes below a few percent at the shortest distances.

 In addition to phonons and plasmon loss, there are several potential sources of error in both the C-Phase gate and the entanglement generation: (i) disorder in the position of the dopants, (ii) enhanced plasmon emission to the excited $1s(T_2/E)$ valley states, and (iii) dephasing and decay of the ground state spins during gate operation.  
(i) The precision with which the dopant atoms must be placed with respect to the nanoribbon and each other is determined primarily by  Eq.~(1) (to achieve sub-radiance their distance from each other must be a half integer multiple of the plasmon wavelength, but this can be tuned with the Fermi energy).   It follows that the position error of the two dopants should be less than $ f/ 2 k$ relative to the nanoribbon.  Since $k = W/\eta^{-1}(x)$, this implies that for $W \approx 50$ nm and $f \lessapprox 0.1$ the donor placement precision should be  $\lessapprox5$ nm for the value of $\eta^{-1}\approx 0.6$ used in Fig.\ 3(b).  Such precision is achievable with current technology. (ii)  In addition, to the enhanced plasmon decay to the ground state, there will also be enhanced  decay to the $1s(T_2/E)$ states which cannot be made sub-radiant.  However, when the system is optimized for the ground state transition, the enhanced emission to the other $1s$ states scales as $\gamma_p (\omega_{T,E}/\omega_d)^9$.  For Si:P(Bi) the ratio $(\omega_{T,E}/\omega_d)^9 \approx 10^{-2}(10^{-4})$ so for P it is a small correction and for Bi it is negligible.  (iii)  For Si:P(Bi) qubits electron spin coherence times as long as a  few ms have been observed \cite{Zwanenburg13}.  This sets a lower limit on the gate speed which is proportional to $\Omega^2/\gamma$.  At 1 $e\cdot$pm, the dipole moment of the $D^0X$ transition is extremely weak; thus, a strong laser is required to achieve sufficiently fast gates, e.g., at a laser intensity of 10 W/cm$^2$ the gate time will be roughly 10 $\mu$s, which is slow, but still much faster than the decoherence rate of the electron spins.  A key advantage of the group-V donor qubits in Si is that, once the gate or entanglement operation has been performed, the electron spin states can be mapped to the donor nuclear spin, where coherence times over an hour have been demonstrated \cite{Saeedi13}. 
 
 We have shown that mid-infrared graphene plasmonics may be a powerful resource for optical manipulation and inducing interactions of group-V donor spin qubits in Si.   
In addition to the  applications described here, such a system could be a powerful resource for photonic applications, e.g., an electrically pumped THz laser \cite{Apalkov13}.  
The main experimental challenges will be combining high-quality graphene nanoribbons and the precise placement of  impurities in the Si lattice.    The already dramatic progress along these fronts suggests that graphene plasmons may enable a new route towards spin based  metrology and quantum computation in Si. 

\emph{Acknowledgments --} We thank G.\ Wolfowicz, J.\ J.\ L.\ Morton, M.\ Keller and D.\ Newell for helpful discussions.  Funding is provided by NIST and the  NSF Physics Frontier at the JQI.

\bibliographystyle{../apsrev-nourl}

\bibliography{../SiVG_bib}

\setcounter{equation}{0}
\renewcommand{\theequation}{S\arabic{equation}}
\setcounter{figure}{0}
\renewcommand{\thefigure}{S\arabic{figure}}

\appendix
\newpage
\section{Supplemental Material}

\begin{figure*}[t]
\begin{center}
\includegraphics[width=0.95 \textwidth]{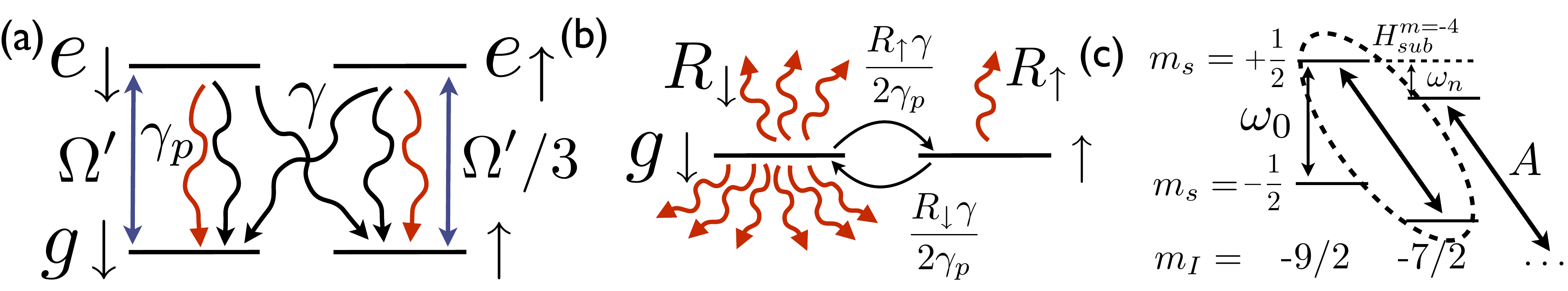}
\label{fig:SiV_supp_Fig1}
\caption{(a) Reduced level diagram for the $\sigma_-\sigma_+$-excitation scheme.  (b) Effective ground state evolution for the single-shot readout.  Nine times as many plasmons are emitted in the spin-down state.  (c) Hyperfine states for Si:Bi.  There is an electron(nuclear) Zeeman energy $\omega_{0(n) }$ and a Hyperfine coupling between pairs of states in the submanifolds.  For the protocols in Si:Bi we use the pseudo-spin-1/2 subspace of $\ket{g,\downarrow}=\ket{m_s=-1/2,m_I=-9/2}$ and $\ket{g,\uparrow}=\ket{+,-4}$. At large magnetic field these two states correspond to up/down electron spin and the fully polarized nuclear spin $m_I=-9/2$.}
\end{center}
\end{figure*}

\emph{Device Properties --}
The graphene could be encapsulated between two thin layers of hexagonal-BN (h-BN), while the doping of the nanoribbon is  controlled by either direct chemical doping or a metal top gate.  Room temperature mobilities approaching 10$^6$ cm$^2$/Vs have been demonstrated for bulk graphene surrounded by hBN \cite{Mayorov11,Elias11}, which would allow mid-infrared plasmons to coherently propagate for several microns \cite{Jablan09}.  The $1s$ states of the donor have a six-fold valley degeneracy, which splits into a single orbital ground state $A_1$ and five excited states lying 10-30 meV higher in energy: a doublet $E$ and triplet $T_2$.   Due to their large overlap with the donor nucleus they have a relatively large spin-orbit coupling [$\sim0.02(1)$ meV in Si:P(Bi)].   The excited state transitions from $1s(A_1)$ to $2p$ states lie in the mid-infrared  between 30-70 meV.  The $2p$ states primarily decay via phonons in a two step process through the $1s(T_2/E)$ valley states, but  have long lifetimes: observed as long as 250 ps \cite{Vinh08} and predicted to be as long as 1.3 ns \cite{Tyuterev10}.  The magnitude of the dipole moment for the $1s$ to $2p$ transition is $\abs{\bra{e_s}\mu_I\ket{g,s}} \approx 0.3~e\cdot$nm.  In addition to the valley and spin fine structure, there is a hyperfine interaction with the donor nucleus, which has been used to achieve long term quantum memory in these systems \cite{Mohammady12,Wolfowicz13}; however,  for all the donors except Bi (see below), the hyperfine interaction can be neglected at  moderate magnetic fields  $\gtrsim 10$ mT.

The spin-$3/2$ character of the $D^0X$ state has been measured and used for spin state preparation and detection in ensembles of Si:P and Si:Bi \cite{Yang06,Sekiguchi10}.
The $D^0X$ states lie $1.1$ eV above the $1s(A_1)$ ground state and have a lifetime of 272 ns, limited by Auger recombination.  In principle, a  lambda transition from the ground state to the to the 2$p$ states, through the $D^0X$ states, is allowed, but it is parity non-conserving, which is only weakly broken in this system.  Instead we propose to hybridize the $2s$ states with the $2p$ states through application of an external stress, to make them degenerate, and an external electric field to mix them.  This requires a pressure of $\sim 25$ MP along one of Si principal axes \cite{Ramdas81}. To apply suitable electric fields, one could either use an external gate or the nanoribbon itself, which will induce a 2s-2p splitting as large as $\approx 1$ meV for doping levels of $n_e\approx 10^{12}$ cm$^{-2}$.

\emph{Fidelity of Single-Shot Readout} -- 
To achieve single-shot readout we use the $\sigma_-\sigma_+$-excitation scheme discussed in the main text and shown in Fig.\ S1(a).  The Hamiltonian for the Raman Rabi driving field takes the form
\begin{align}
H_{c}&=\Omega'(t)\bigg( \frac{1}{3} \ket{e_\uparrow}\bra{g,\uparrow} + \ket{e_\downarrow}\bra{g,\downarrow} + h.c. \bigg)
\end{align}
The factor of $1/3$ follows from the ratio of Clebsch-Gordon coefficients in the two excitation pathways  through the spin-3/2 exciton state. 
When the impurity is driven resonantly below saturation,  the fluorescence distribution of the emitted plasmons follows a Poisson distribution  \cite{Kimble76}, which is different for the two spin states 
\be
p_s(n,t)=\frac{(R_s t)^n}{n!} e^{-R t}
\ee
where $R_\downarrow=\frac{2 \Omega'^2}{(\gamma+\gamma_p)^2}  \gamma_p$ and $R_\uparrow = \frac{2 \Omega'^2}{9(\gamma+\gamma_p)^2} \gamma_p$.    When the excited state emits a phonon instead of a photon it decays through the $1s(T_2/E)$ valley states where it can flip its spin due to the large spin orbit coupling in these states \cite{Ramdas81}.  As a worst case scenario, we assume that each time a phonon decay occurs, there is an electron spin flip with probability 1/2.  This leads to the  evolution for the ground state spin populations $n_{\uparrow/\downarrow}$
\be
\dot{n}_\uparrow = - \frac{\gamma}{2\gamma_p} R_\uparrow\, n_\uparrow + \frac{\gamma}{2 \gamma_p} R_\downarrow\, n_\downarrow 
\ee
and $n_\downarrow \approx 1- n_\uparrow$ since we are far below saturation.   The dynamics are illustrated in Fig.\ S1(b) The emitted  plasmon number distribution $P(n,t)$ is
\begin{align}
P(n,t)&=n_\uparrow(t) p_\uparrow(n,t)+n_\downarrow(t) p_\downarrow(n,t), \\
n_\uparrow(t)& = n_\uparrow(0) e^{- \gamma (R_\uparrow+R_\downarrow) t/\gamma_p}+ \frac{1}{10}\big[1-e^{- \gamma (R_\uparrow+R_\downarrow) t/\gamma_p} \big].
\end{align}
In Fig.\ 2(b) of the main text we show an example of the distribution for the two initial conditions $p_\uparrow=1$ and 0. The distribution for the two initial spin states will be disjoint when $\gamma_p/\gamma \gg R_\downarrow \, t \gg 1$ which guarantees that many plasmons are emitted before the spin is depolarized.  The optimum occurs when $R_\downarrow \, t \sim \sqrt{2\gamma_p/\gamma}$.

\emph{Hyperfine Interaction in Si:Bi} --
The Hamiltonian for the ground state spin interacting with the donor nucleus is
\be
H= \omega_0 s_z-\omega_n I_z + A\, \bm{s}\cdot \bm{I}
\ee
where $I_z$ is the donor nuclear spin operator, $I=9/2$ in the case of Si:Bi, $\omega_{0(n)}$ is the Zeeman energy of the electron(nuclear) spin, and $A$ is the hyperfine coupling constant.  For Si:Bi the hyperfine coupling is rather large $A/\hbar = 1.47$ GHz.  To find the eigenstates, we first  notice that $H$ has the two decoupled eigenstates $\ket{m_s=\pm1/2,m_I=\pm I}$, where $m=m_s+m_I$ is the total z-angular momentum.  The other eigenstates are formed from superpositions of pairs of states $\ket{m_s= \pm1/2,m_I=m\mp 1/2}$ with  $\abs{m}<I+1/2$ as shown in Fig.\ S1(c).  We can define Pauli operators within each of these two-state submanifolds, then we can rewrite the Hamiltonian for these submanifolds as \cite{Mohammady12}
\begin{align}
H^m_{sub}& = \frac{\Delta_m}{2} \sigma_z+ \frac{\Omega_m}{2} \sigma_x- \epsilon_m \\
\Delta_m & = A\, m+ \omega_0+\omega_n \\
\Omega_m &=A \sqrt{I(I+1)-m^2 }\\
\epsilon_m &=\frac{A}{4}+\omega_n m
\end{align}
Defining $\theta_m = \tan^{-1}(\Omega_m/\Delta_m)$, the eigenstates are 
\be
\begin{split}
\ket{\pm, m} &= \cos(\theta_m/2) \ket{m_s=\pm\frac{1}{2},m_I=m\mp \frac{1}{2} } \\
&\pm \sin(\theta_m/2) \ket{m_s=\mp \frac{1}{2},m_I=m\pm \frac{1}{2} }
\end{split}
\ee
To apply the results discussed in the main text to case of Si:Bi we work in the pseudo-spin 1/2 subspace with $\ket{g,\downarrow}=\ket{m_s=-1/2,m_I=-9/2}$ and $\ket{g,\uparrow}=\ket{+,-4}$.  At large magnetic field these two states correspond to up/down electron spin and the fully polarized nuclear spin $m_z=-9/2$. The $\ket{g,\downarrow}$ state can be prepared by optically pumping with the $\sigma_-\pi$-excitation scheme.  In addition, we note that both the two-qubit C-Phase gate and the entanglement preparation carry through unaltered in this pseudo-spin-1/2 subspace.

\emph{Master Equation --}
When the propagation time between impurities $r/v_g$ is much faster than the other time scales we can integrate out the plasmon modes in a Markov approximation and the super-radiance can be clearly seen in the master equation for the density matrix $\rho$ of the two impurities
 \begin{align} \nonumber
\dot{\rho} &= -\frac{i \omega_{d}}{2} \comm{\sigma_1^z+\sigma_2^z}{\rho}- \frac{i}{2} v \comm{(\sigma_1^+\sigma_2^- +\sigma_1^-\sigma_2^+)}{\rho} \\  \label{eqn:me}
&- \gamma_c \mathcal{D}[\sigma_1^-+ \alpha \,\sigma_2^-]\rho  - {\delta \gamma}( \mathcal{D}[\sigma_1^-] + \mathcal{D}[\sigma_2^-]) \rho \\
v &=  \gamma_p\, \sin k r \, e^{-r/L_p}, ~~~~~~ \gamma_c =  \gamma_p \,\abs{\cos kr} \, e^{-r/L_p}  \\
\delta \gamma& = \gamma+\gamma_p - \gamma_c,~~~~ L_p^{-1} = \frac{e\,  v_F^2}{2 \mu\,v_g \omega_F}
 \end{align}
 where $\mathcal{D}[c] \rho =1/2 \{c^\dagger c,\rho\}-c \rho c^\dagger$ and $\alpha= \textrm{sign} (\cos kr)$.  We have included the losses in the plasmon propagation in terms of  the plasmon propagation length $L_p$ defined in the main text. 
 

%

\end{document}